# Optically induced coherent transport far above $T_c$ in underdoped $YBa_2Cu_3O_{6+\delta}$

S. Kaiser[1*], D. Nicoletti[1*], C.R. Hunt[1,4*], W. Hu[1], I. Gierz[1], H.Y. Liu[1], M. Le Tacon[2],

T. Loew[2], D. Haug[2], B. Keimer[2], and A. Cavalleri[1,3]

[1] *Max Planck Institute for the Structure and Dynamics of Matter, Hamburg, Germany*
[2] *Max Planck Institute for Solid State Research, Stuttgart, Germany*
[3] *Department of Physics, Oxford University, Clarendon Laboratory, Oxford, United Kingdom*
[4] *Department of Physics, University of Illinois at Urbana-Champaign, Urbana, Illinois, USA*

**We report on a photo-induced transient state of $YBa_2Cu_2O_{6+\delta}$ in which transport perpendicular to the Cu-O planes becomes highly coherent. This effect is achieved by excitation with mid-infrared optical pulses, tuned to the resonant frequency of apical oxygen vibrations, which modulate both lattice and electronic properties. Below the superconducting transition temperature $T_c$, the equilibrium signatures of superconducting interlayer coupling are enhanced. Most strikingly, the optical excitation induces a new reflectivity edge at higher frequency than the equilibrium Josephson plasma resonance, with a concomitant enhancement of the low frequency imaginary conductivity $\sigma_2(\omega)$. Above $T_c$, the incoherent equilibrium conductivity becomes highly coherent, with the appearance of a reflectivity edge and a positive $\sigma_2(\omega)$ that increases with decreasing frequency. These features are observed up to room temperature in $YBa_2Cu_2O_{6.45}$ and $YBa_2Cu_2O_{6.5}$. The data above $T_c$ can be fitted by hypothesizing that the light re-establishes a transient superconducting state over only a fraction of the solid, with a lifetime of a few picoseconds. Non-superconducting transport could also explain these observations, although one would have to assume transient carrier mobilities near $10^4$ cm$^2$/sec at 100 K, with a density of charge carriers similar to the below $T_c$ superfluid density.**

Our results are indicative of highly unconventional non-equilibrium physics and open new prospects for optical control of complex solids.

Doped cuprates retain important properties of the superconducting state above the transition temperature $T_c$[1,2,3]. Charge transport in the normal state is highly incoherent, with low carrier mobilities[3]. Fluctuations of the order-parameter phase[4,5,6] and the emergence of competing orders[7,8] arise as coherence is lost when the temperature is raised above $T_c$. Magnetic fields affect this interplay, quenching superconducting coherence in favor of the competing phase[9,10]. In this work, we investigate the effect of optical excitation, with the goal of enhancing coherence and achieving the opposite effect of a magnetic field.

In the past, optical excitation has been applied almost exclusively at visible or near infrared wavelengths, revealing information on the relaxation of hot incoherent quasiparticles back into the Cooper-pair condensate[11,12,13]. More recently, mode selective optical deformation of the crystal lattice[14,15,16] has been used to affect the balance between competing phases on lower energy scales, transforming an insulating striped phase into a transient superconductor[17].

In the case of underdoped $YBa_2Cu_3O_{6+\delta}$, much work has been dedicated to understanding the properties of the normal state, unveiling a complex interplay of charge[18,19,20] and spin[21] order and possible above-$T_c$ coherence[22,23]. These observations suggest that application of appropriate stimulation may re-establish superconducting order[24].

Three different underdoped compounds were studied in our experiments, $YBa_2Cu_3O_{6.45}$ (YBCO 6.45), $YBa_2Cu_3O_{6.5}$ (YBCO 6.5) and $YBa_2Cu_3O_{6.6}$ (YBCO 6.6), with corresponding hole doping levels of 7%, 9%, and 12%. Crystals of typical dimensions 2 x 2 x 1 mm$^3$ were grown in Y-stabilized zirconium crucibles[25]. The hole doping of the Cu-O planes was adjusted by controlling the oxygen content of the Cu-O chain layer $\delta$ by annealing in flowing $O_2$ and subsequent rapid quenching. The $T_c$ values ($T_c$ = 35 K for YBCO 6.45, $T_c$ = 50 K for YBCO 6.5, and $T_c$ = 62 K for YBCO 6.6) were determined by dc magnetization measurements in a SQUID, as discussed in the supplementary section.

For the purposes of the following discussion, it is important to note that equilibrium superconductors display a characteristic frequency dependent conductivity $\sigma_1(\omega)+i\sigma_2(\omega)$, with a zero-frequency delta function in its real part $\sigma_1(\omega)$ and a positive imaginary part $\sigma_2(\omega)$ that diverges at low frequency as $1/\omega$. This frequency dependent conductivity is clearly distinct from that of a Drude metal, for which charge carriers have a finite scattering time $\tau_s$ of only a few femtoseconds ($10^{-15}$ sec). The real part of the conductivity is then constant for all frequencies smaller than the scattering rate $1/\tau_s$ ($\sim$ tens or hundreds of THz) and vanishes above it. The imaginary part $\sigma_2(\omega)$ is instead peaked at $1/\tau_s$, and tends to zero both at lower and higher frequencies.

For the case of layered cuprates, the difference between coherent and incoherent transport is underscored by further features of the optical properties. A zero crossing of the real part of the dielectric permittivity $\varepsilon_1(\omega)$ and a peak in the loss function $-\mathrm{Im}(1/(\varepsilon_1(\omega)+i\,\varepsilon_2(\omega))$ result in the appearance of a characteristic reflectivity edge at $\omega_J$, the so-called Josephson Plasma resonance. At $\omega_J$, self-sustaining oscillations of Cooper-pairs tunnel between capacitively coupled planes, giving rise to these features. Above the superconducting transition temperature the reflectivity becomes featureless around $\omega_J$, dominated by incoherent scattering at rates faster than the tunneling oscillations.

In the experiments reported here, changes in the equilibrium optical properties were induced by excitation with mid-infrared optical pulses of $\sim$300 fs duration, polarized along the *c* direction and tuned to the 20 ± 3 THz frequency (~15-μm, 670 cm$^{-1}$, 83 meV, ± 15%) of the infrared-active distortion of figure 1, which modulates the apical oxygen positions. The excitation pulses were generated by difference-frequency mixing in an optical parametric amplifier and focused onto the samples with a maximum fluence of 4 mJ/cm$^2$, corresponding to peak electric fields up to ~3 MV/cm. At this excitation level the vibrational mode was excited by several percent of the equilibrium bond distance.

The equilibrium and transient optical properties were probed in reflection between 0.5 and 2.5 THz[26,27]. Single-cycle THz pulses were generated by optical rectification of a near-infrared (800-nm wavelength) femtosecond pulse in a ZnTe crystal. The probe pulses were focused onto the YBCO crystals with polarization perpendicular to the superconducting planes (*c* axis), and were electro-optically sampled after reflection by a second near-infrared (800-nm wavelength) pulse in a second ZnTe crystal.

Figure 2 summarizes the "raw" data obtained in the superconducting state (10 K base temperature). The equilibrium optical properties show reflectivity edges at characteristic frequencies at $\omega_J \sim 20$ cm$^{-1}$ for YBCO6.45, $\omega_J \sim 30$ cm$^{-1}$ for YBCO 6.5, and $\omega_J \sim 60$ cm$^{-1}$ THz for YBCO 6.6 (figure 2b).

After excitation with the optical pump, the changes in the reflected electric field were recorded, as displayed in panels (a1)-(a3). Note that a clear oscillatory response, containing a dominant frequency component is observed in these differential changes, indicating that a single resonance is being created or modified. From these curves, the differential electric field $\Delta E_R(t, \tau)$ and the stationary reflected electric field $E_R(t)$ were independently Fourier transformed to obtain the complex-valued, frequency dependent $\Delta \tilde{E}_R(\omega)$ and $\tilde{E}_R(\omega)$. With knowledge of the equilibrium optical properties of the material the full complex optical response can be evaluated without using Kramers-Kronig relations.

The lower panel (b2) shows the corresponding reflectivity changes measured 0.8 ps after the photo-excitation[28], where the signal was maximum. For each doping level, we observed the strongest change around $\omega_J$, with a reduction below the resonance and an enhancement above, indicating a blue shift of the plasma edge.

By analyzing the amplitude and phase of the transient reflectivity changes, taking into account the equilibrium optical properties and the mismatch between the penetration depths of mid-IR pump and the THz-probe beam (see Supplementary material), we determine the transient optical conductivity $\sigma_1(\omega) + i\sigma_2(\omega)$ in the photo-excited volume alone. Figure 3 displays the results of this analysis for YBCO 6.45, 6.5 and 6.6 in which the equilibrium optical properties are presented as gray lines, with the top panels indicating the real and imaginary part of the optical condutivities $\sigma_1(\omega)$ and $\sigma_2(\omega)$, respectively.

The response after photo-excitation with the mid-IR light pulses is presented with colored lines. The blue, red, and green lines show the optical properties in the photostimulated top layer at a pump-probe time delay of 0.8 ps after the photo-excitation in YBCO 6.45, 6.5, and 6.6. The real part of the optical conductivity is only slightly enhanced, likely due to moderate quasi-particle excitation. Crucially, the imaginary part shows a clear enhancement of the positive and diverging response. For a superconductor, the strength of this low frequency divergence reflects an enhancement in the superfluid density, identified at equiliubrium as proportional to $\lim_{\omega\to 0} \omega\sigma_2$. This enhancement of the low frequency $\sigma_2$ is opposite to the effect observed in photo-doping experiments that break the superconducting condensate and in which the diverging $\sigma_2$ vanishes and a strong quasiparticle excitation sets in[26,27].

In Figure 3(a3)-(c3) we show also the transient photoinduced differential conductivity $\Delta\sigma_2$. These further emphasize the strong enhancement of the low frequency $\sigma_2$ and therefore increase of the superfluid density. We find a stronger enhancement with a steeper divergence to lower frequencies when increasing the doping level.

Next we turn to base temperatures above the critical temperature $T_c$, where equilibrium transport is highly incoherent and the reflectivity featureless. Figure 4 summarizes the 'raw' measured data for photo-excited YBCO 6.45 ($T_c = 35$ K). Panel (a) displays the measured changes in the transient reflected THz-field at 0.8 ps after the mid-IR excitation pulse. The corresponding frequency resolved change in above-$T_c$ reflectivity is reported in panel (b), showing the appearance a clear reflectivity edge of about 2% already in the raw data. The time dependent response of the transient state is shown in panel (b) for the peak of the transient field changes. Note that the lifetime of the light induced state involves two decay timescales of ~0.5 ps and ~5-7 ps. We note that the narrow spectral width of the photo-induced edge of about ~20 cm$^{-1}$ requires a lifetime of the state of more than 2 ps. We attribute the short timescale to the quasi-particle dynamics during the decay of the photoexcited state, which is discussed in more detail in our quantitative analysis. In contrast, the lifetime of the below $T_c$ response, shown as a dashed blue line, exhibits basically a single exponential decay and lasts more than 10 ps.

The size of this reflectivity edge is still about 1% when the temperature is increased up to 330 K as reported in the temperature dependence in panel (d).

As done for the below $T_c$ data, we analyze the amplitude and phase of the measured reflectivity transient taking into account the pump-probe penetration depth mismatch, and we determine the complex optical conductivity $\sigma_1(\omega) + i\sigma_2(\omega)$ of the photo-excited volume. Figure 5 displays the results of this analysis for YBCO 6.45 at three temperatures (100 K, 200 K, 330 K). A virtually unchanged real part of the conductivity $\sigma_1(\omega)$ (panels 5(a1)-5(c1)) is accompanied by an increase and a change of slope in the imaginary part $\sigma_2(\omega)$ (panels 5(a2)-(c2)), which becomes positive and increases for decreasing $\omega$ (blue points).

Most strikingly, in panels 5(a3)-(c3) the corresponding change in imaginary conductivity $\sigma_2(\omega)$ (difference between the two curves in panels 5(a2)-(c2)) displays an approximate $1/\omega$ divergence over this frequency range.

From the optical properties extracted above, we also calculate the reflectivity changes that one would observe if the entire probed depth was excited (panels 5(a4)-(c4)), yielding a rescaled edge with respect to that already reported in figure 4(b). The corresponding changes in the energy loss function $\Delta(-\text{Im}\,1/\epsilon)$ are presented in panels 5(a5)-(c5) peaking at the frequency of the photo-induced reflectivity edge. The fluence and wavelength dependence of the photo-induced effects are shown in Figure 6. We find them linearly increasing with the electric field of the mid-IR excitation pulses (Fig. 6(a)) and resonant with the phonon-absorption of the IR active apical oxygen mode at 15 μm (Fig. 6(b)).

Figures 7 and 8 show the corresponding temperature dependence of the optical response above $T_c$ for YBCO 6.5 and YBCO 6.6. We find a similar transient response as that of YBCO 6.45, with an enhancement of the low frequency $\sigma_2(\omega)$, the appearance of a plasma-edge in the reflectivity and a peak in the photo-induced loss function. However, when increasing doping an increasing response in the real part of the optical conductivity $\sigma_1(\omega)$ is observed, suggesting that some of the pump energy is incoherently exciting quasi-particles. With increasing doping level the photo-induced effect becomes stronger. The photo-induced plasma edge shifts to the blue, tracking the blue shift in the Josephson plasma edge observed in equilibrium below $T_c$. However, the temperature scale of the transient state does not track equilibrium superconductivity, instead the size of the effect drops more rapidly with temperature at higher dopings. In YBCO 6.6 the effect has entirely disappeared by 300 K, as shown in figure 8(c). The transient reflectivity is flat and $\Delta\sigma_2(\omega)$ is negative.

To quantitatively fit the temperature dependent data, we first note that the reflectivity edge does not change in frequency with increasing temperature, in apparent contradiction with the decrease in $\Delta\sigma_2(\omega)$. This is interpreted as evidence for a state in which only a fraction of the material is transformed, embedded in a medium that remains incoherent. By assuming a high mobility state with constant local carrier density, but distributed in progressively sparser domains within the unperturbed normal state, these optical properties can be fitted very closely.

A quantitative fit to the data was obtained by applying the Bruggeman[29,30] effective dielectric function $\tilde{\varepsilon}_E(\omega)$ for an inhomogeneous medium: $f\frac{\tilde{\varepsilon}_{HM}(\omega)-\tilde{\varepsilon}_E(\omega)}{\tilde{\varepsilon}_{HM}(\omega)+2\tilde{\varepsilon}_E(\omega)} + (1-f)\frac{\tilde{\varepsilon}_{NS}(\omega)-\tilde{\varepsilon}_E(\omega)}{\tilde{\varepsilon}_{NS}(\omega)+2\tilde{\varepsilon}_E(\omega)} = 0$. We considered a mixture of a high-mobility conductor with dielectric function $\tilde{\varepsilon}_{HM}(\omega)$ of volume fraction $f$, which contains the photo-induced plasma edge, and of normal-state, unperturbed YBCO with the experimentally determined equilibrium $\tilde{\varepsilon}_{NS}(\omega)$[34]. The experimental observations are fit by letting only the plasma frequency $\omega_P$ and the transformed volume fraction $f$ vary as free parameters[31], yielding curves that closely match the experimental observations (see black curves in figures 5, 7 and 8).

In figure 9, the key observations are summarized. We show an increase in $\sigma_2(\omega)$ and the appearance of a reflectivity edge and the corresponding loss function peak, are shown for YBCO 6.45, YBCO 6.5 and YBCO 6.6 at 100 K for a direct comparison. We identify the increase of the inductive response in the differential optical conductivity $\Delta\sigma_2(\omega)$ and in the photo-induced plasma edge. The plasma frequency and the corresponding loss function peak increases with increasing doping level. Note once again how the photo-induced reflectivity edge and loss function peak are still observed at room temperature in YBCO 6.45 and YBCO 6.5, but they are no longer present in YBCO 6.6.

The results of the fits are summarized in figure 10. In panels 10(a1) and 10(a2), we show the optical conductivity of the normal state $\tilde{\varepsilon}_{NS}(\omega)$ (as extracted from literature data[34]) and of the transformed regions, $\tilde{\varepsilon}_{HM}(\omega)$. Most conservatively, these optical conductivities could be fitted by assuming the transformed volume consists of a gas of non-interacting Drude electrons with conductivity $\tilde{\sigma}(\omega) = \omega_p^2/4\pi(1/\tau - i\omega)$ [32] with the experimentally determined plasma frequency $\omega_p^2 = (4\pi n e^2)/m_e$ and a scattering time $\tau$ = 7 ps, corresponding to the lifetime of the photo-induced state, as extracted from the measurement in Figure 4(b). With this model we can fit all temperatures and doping levels, and over the full range of measured fluences (see supplementary information). Fits are shown as black lines in figures 5, 7, 8 and 9. From the simple Drude model we extract a DC mobility $\mu = e\tau/m_e \sim 12000$ cm$^2$/V·s. We stress that such high mobility is highly unusual for c-axis transport in oxides, comparing favorably even with modulation-doped GaAs heterostructures in the same temperature range. Carrier mass renormalization would lower the effective mobility[2,3].

A transient superconducting state with lifetime $\tau_L \sim 7$ ps would exhibit the same ac optical properties as a Drude conductor with scattering time $\tau_S = \tau_L$. Since the minimum frequency measured here is greater than $1/\tau_L$, the transient state can be equally well described assuming the transformed volume is an equilibrium superconductor, ie $\tau_L \rightarrow \infty$ (red lines in same figures), but could not be fitted with scattering times below the lifetime of the state. At room temperature, this corresponds to effective scattering times of about 2 ps, which would reduce the extracted mobility by a factor of about 3.5 compared to the above.

The value of the plasma frequency $\omega_p$ used to fit the data for each doping value is almost independent of the base temperature, whereas the volume fraction $f$ extracted with the fit decreases with increasing temperature (see below). In panels 10(b1)-10(b3) we display the temperature dependence of $f$ for the three different doping levels measured. A temperature scale T' for the light induced high mobility phase is established by fitting the temperature dependent volume fraction with an empirical mean-field law of the type $\propto \sqrt{1 - T/T'}$. The fitted T' values (T'$_{6.45}$=370±25 K, T'$_{6.5}$=330±10 K, and T'$_{6.6}$=160±20 K) are visually reported in the phase diagram figure 11 where the blue shaded area indicates the region where we can photo-induce a coherent high mobility state with scattering times always as long as, or even longer than, the lifetime of the transient state.

The interpretation of a high-mobility Drude conductor is in our view unlikely. For example, one cannot invoke a gas of photo-excited carriers with unusually long scattering times. The fraction of material that was switched is a function of laser field, but the position of the edge exhibited no fluence dependence and was independent of the number of absorbed photons (see supplementary information). For a gas of carriers photo-excited above a gap, even in the absence of damping, one would expect the position of the edge to shift to the blue following the carrier density as $n^{1/2}$.

An alternative effect that could give rise to such high mobility transport may be conduction by a non-commensurate sliding one-dimensional charge density wave[33], which could become de-pinned when the lattice is modulated and pin again a few picoseconds after excitation. Indeed, as charge density waves have been observed in this doping range, this possibility should be considered as a possible explanation.

More than one observation supports an interpretation for the above $T_c$ data based on transient superconducting coherence. First of all one can compare all the light induced changes reported here to the observations made for the transient state induced below $T_c$.

As discussed in Figure 3 a clear photo-induced enhancement of the superconducting state is seen below $T_c$, as underscored by a blue shift of the Josephson plasma edge and the strong increase in the divergence of the low frequency $\sigma_2(\omega)$. Furthermore the photo-induced change $\Delta\sigma_2(\omega)$ in imaginary conductivity above $T_c$ displayed in panels 5(a3)-5(c3) tracks very well the superconducting component of the equilibrium inductive conductivity below $T_c$, $\Delta\sigma_2(\omega) = \sigma_2(\omega, 10\,\text{K}) - \sigma_2(\omega, 60\,\text{K})$, which is indicated by a grey dashed curve in figure 5(a3). This similarity is further underscored by the observation that the position at which the plasma edge is being generated is very close to the equilibrium Josephson Plasma Resonance. Also, the frequency of the light-induced edge shifts to higher frequencies for increasing static doping, in analogy with the equilibrium Josephson Plasma Resonance[34] (see Figure 2(b1) and supplementary information).

To identify the physical mechanisms that may cause transient superconductivity, we first note how photo-induced redistribution of quasi-particles[35,36], which was shown in the past to enhance superconductivity at microwave[37,38,39,40,41] or optical[42,43] frequencies is an unlikely explanation. The light-induced state can be created only when the pump wavelength is tuned to the phonon resonance (see figure 6), disappearing at higher frequency. This wavelength dependence is incompatible with charge excitations, which should persist at photon energies higher than the phonon frequency.

Rather, excitation of the lattice may be "melting" an ordered state that competes with superconductivity, as shown in the past for striped cuprates and for other complex materials. In underdoped YBCO, the recent discovery of a Charge Density Wave (CDW) that competes with superconductivity[8] provides an appealing physical framework for this effect. On the other hand, the anti-correlation of the light-induced superconducting volume fraction, $f$, (fig. 10B) with the equilibrium CDW amplitude measured upon doping would argue against a scenario in which light illumination primarily modulates the balance between CDW and superconducting phases. It is interesting to note that $f$ does correlate with the spectral weight of low-energy spin excitations, which is most pronounced in YBCO 6.45 and decreases strongly with both doping[44] and temperature dependence[45].

A second possibility is that the nonlinear coherent phonon excitation[46,47] may be transiently creating a displaced crystal structure with atomic positions more favorable to high temperature superconductivity. Indeed, appropriate displacements of the apical oxygen away from the planes may promote coherence[48,49,50,51,52]. Finally, one should also consider the possibility of a coherent state induced by rapid modulation. As the 20-THz modulation used here occurs at frequencies high compared to plasma excitations between planes, one could envisage a dynamically-stabilized stack of Josephson coupled planes[53]. An instructive analogy can be drawn with other physical systems in which coherence is enhanced by rapid modulation, as observed in classical mechanics[54,55] or in Nuclear Magnetic Resonance[56]. The lifetime of the state (see figure 2) is compatible with the dephasing time of the driven infrared active vibration.

In summary, we have used vibrational excitation in the mid-infrared to induce a transient state with high c-axis mobility, with a carrier scattering time $\tau_s \gg 1$ ps. The optical data are fit by assuming that the high-mobility state occupies a volume fraction that decreases with increasing temperature. Stronger pump fields, impossible in our apparatus, may make it possible to reach the percolation threshold for the high mobility phase at which point one may observe a homogeneous phase, a qualitatively different response and longer lifetimes.

Interpretations based on conduction by normal carriers with anomalously high mobility or by a transient superconducting state fit the data equally well. Our experiments provoke new thinking on the nature of the normal state. We establish a temperature scale $T' \gg T_c$ below which the system becomes a high-mobility conductor. There seems to be a very visual analogy between the onset temperatures of the non-equilibrium perfect conductor and the onset of the preexisting coherence found in optical measuerements[22,23] or more likely even the pseudogap phase (see figure 11). Our fit suggests that high mobility transport or transient superconductivity emerges from progressively sparser regions of the pseudogap, as if only a fraction of the materials were susceptible to being switched[57]. A challenge for future research will be to encapsulate our findings in a description of the physics of coherently-modulated unconventional superconductors.

**Figure 1**

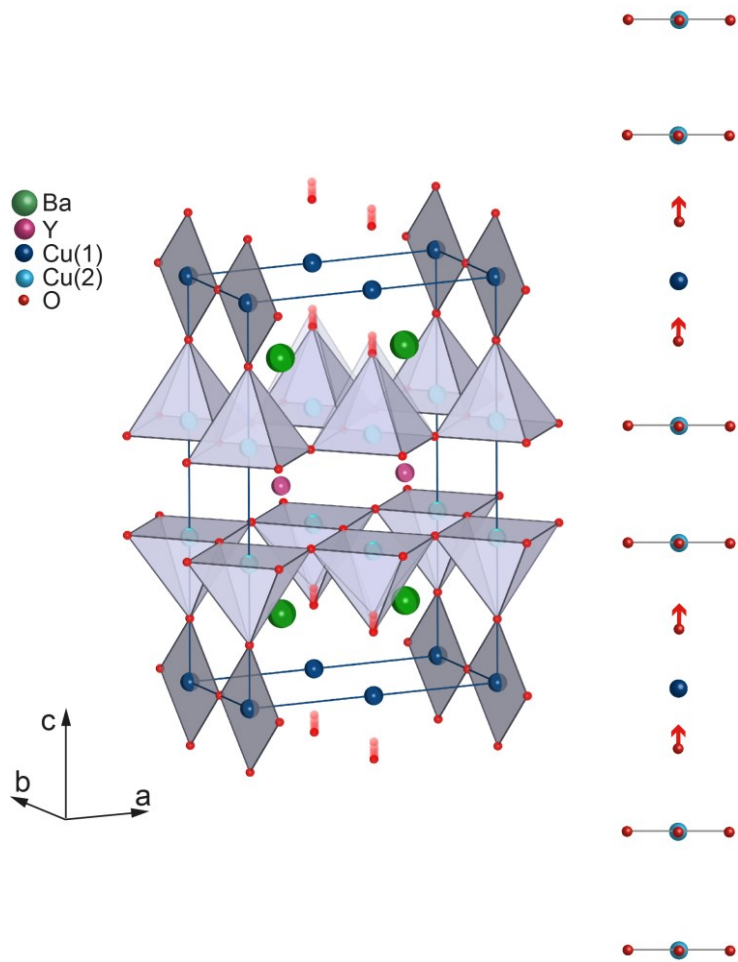

**Figure 1.** Structure of $YBa_2Cu_3O_{6.5}$ and lattice distortion for the 20-THz mode. Two conducting $CuO_2$ planes (Cu(2), and O in the *ab*-plane) are separated by Y atoms (pink) and form a bilayer unit. Ba atoms (green) and the $CuO_4$ ribbons (Cu(1), and O in the *bc*-plane) separate bilayer units[58]. The excitation of the infrared-active $B_{1u}$ mode at 20-THz frequency modulates only the displacement of the apical oxygen atoms along the *c* direction[59].

**Figure 2**

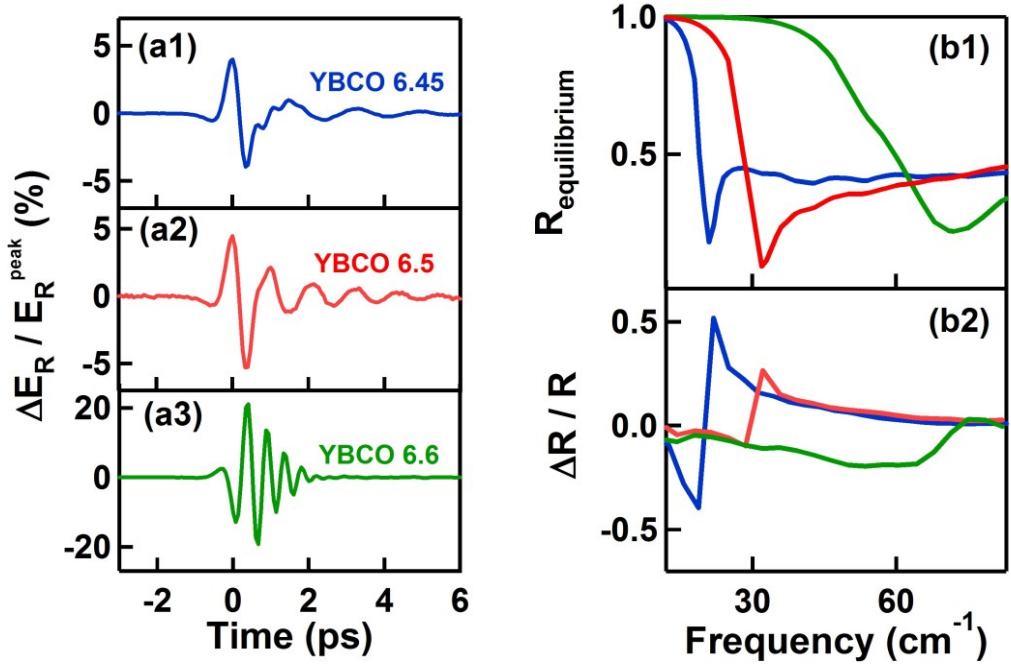

**Figure 2. Below T$_c$ data. (a)** Differential electric-field transient $\Delta E_R(t,\tau)/E_R(t^{peak})$ in YBCO 6.45 (blue), YBCO 6.5 (red) and YBCO 6.6 (green) excited with 20-THz pulses in the superconducting state at 20 K base temperature, measured with THz pulses polarized along the *c*-axis (perpendicular to the superconducting planes). The response was measured at a time delay τ = +0.8 ps after the 20-THz pump. **(b1)** Frequency dependent reflectivities $R_0$ of YBCO 6.45, YBCO 6.5 and YBCO 6.6 measured at equilibrium along the *c*-axis. Reflectivity edges of the Josephson plasma resonance are observed at ω$_{J,6.45}$~ 20 cm$^{-1}$, ω$_{J,6.5}$~ 30 cm$^{-1}$, and ω$_{J,6.6}$~ 60 cm$^{-1}$. **(b2)** Corresponding frequency dependent differential changes in the reflectivities $\Delta R(\omega, 0.8\,\text{ps})/R_0$ measured in Figures a.1-a.3 at positive time delays after the 20-THz pump. A shift of the Josephson plasma resonance is indicated.

**Figure 3**

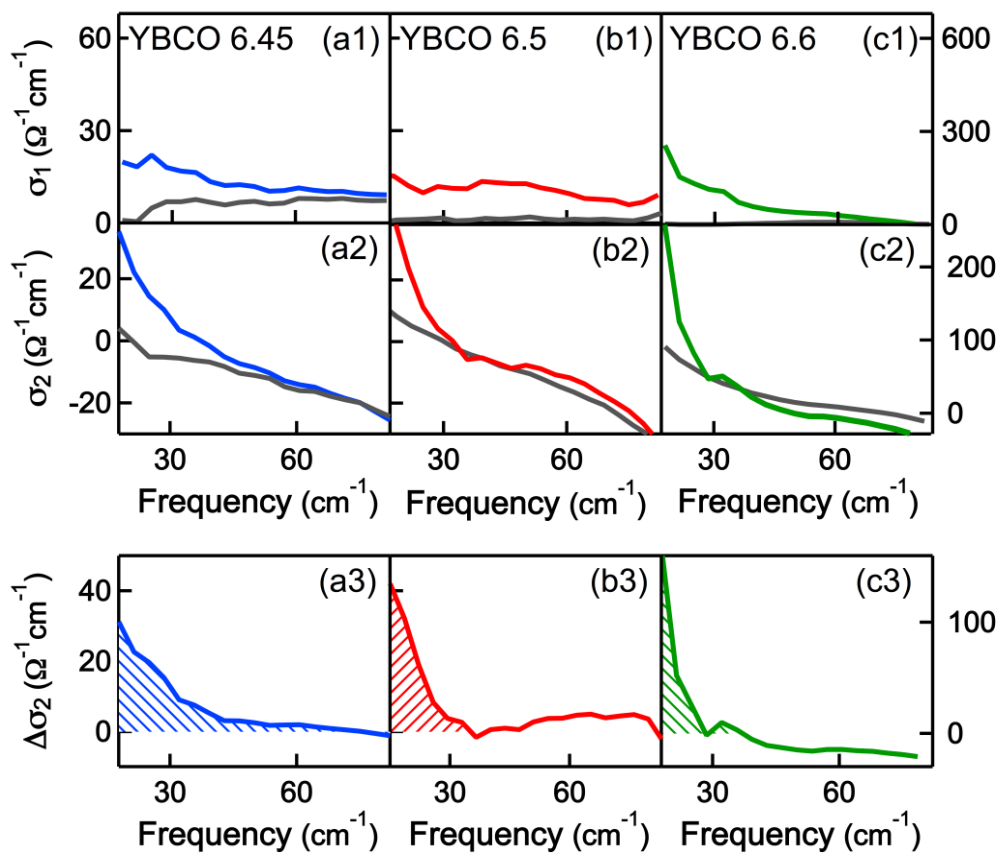

**Figure 3 Below T$_c$ data.** Optical properties of (a) YBCO 6.45, (b) YBCO 6.5, and (c) YBCO 6.6 in the superconducting state at 10 K base temperature. The gray lines describe the optical properties at equilibrium while the colored lines the response in the photo-stimulated state 0.8 ps after the excitation. The upper rows show σ$_1$, the real part, and σ$_2$, the imaginary part, of the complex optical response. The bottom row shows the photo-induced differential optical conductivity Δσ$_2$ between the optical properties in the photo-stimulated and the equilibrium state.

**Figure 4**

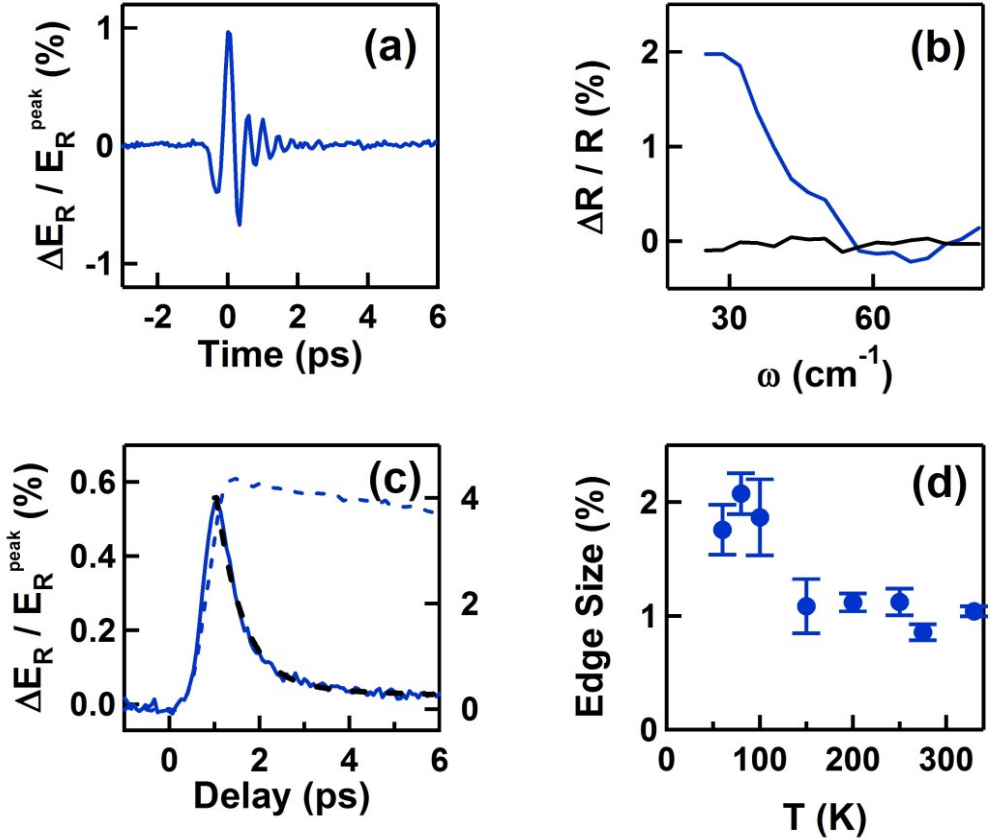

**Figure 4 Above T_c data. (a)** Differential electric field transient $\Delta E_R(t,\tau)/E_R(t^{peak})$ in YBCO 6.45, excited with 20-THz pulses at 100 K base temperature. The response is measured with THz pulses polarized along the *c*-axis (perpendicular to the superconducting planes). The response was measured at a positive time delay ($\tau$ = +0.8 ps) after the 20-THz pump. **(b)** Frequency dependent reflectivity changes measured along the *c*-axis. The black line describes the equilibrium response at negative pump-probe time delays and the blue curve the response at 0.8 ps pump probe time delay after the excitation pulse. A photoinduced reflectivity edge is observed at $\omega_{J,6.45} \sim 43$ cm⁻¹. **(c)** Delay dependence of the peak THz signal $\Delta E_R(t^{peak},\tau)/E_R(t^{peak})$ as measured by scanning the pump-probe delay $\tau$ for YBCO 6.45. $\Delta E_R(t^{peak},\tau)$ is proportional to the frequency integrated effect. For comparison the blue dashed line describes the response below T_c. **(d)** The size of the photo-induced reflectivity edge as function of base temperature.

**Figure 5**

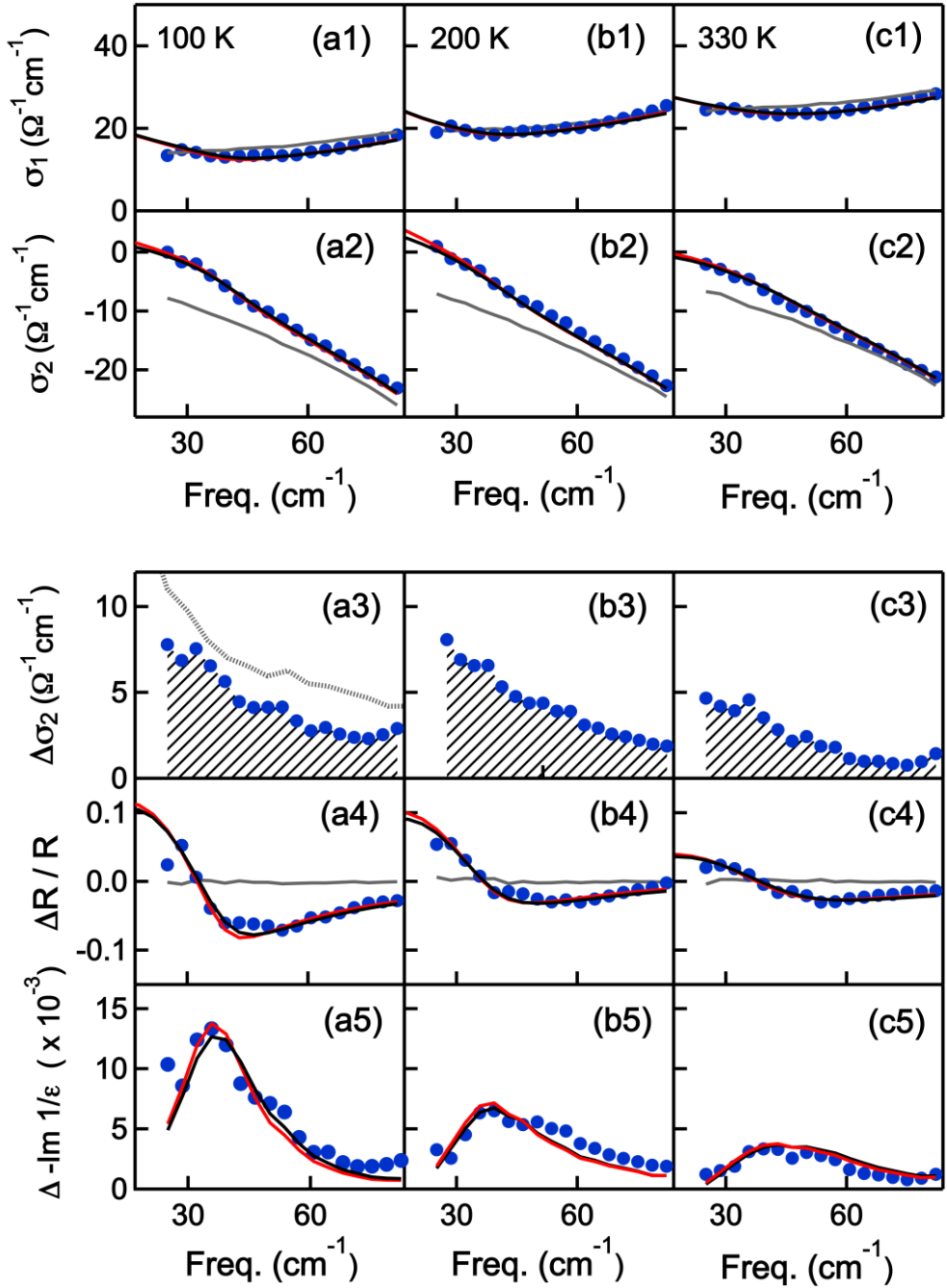

**Figure 5 Optical properties YBCO 6.45 above Tc.** 0.8 ps time delay after excitation at (a) 100 K, (b) 200 K and (c) 330 K. Blue dots: Light induced optical properties. Grey lines: properties of the equilibrium state. Black lines: effective medium fit for a transient conductor with $\tau_s$=7 ps. Red lines: effective medium fit for a perfect conductor with infinite lifetime. **(a1)-(c1)** Real part of the optical conductivity $\sigma_1(\omega)$. **(a2)-(c2)** Imaginary part of the optical conductivity $\sigma_2(\omega)$. **(a3)-(c3)** Differential changes in the imaginary conductivity $\Delta\sigma_2(\omega, 1\text{ ps}) = \sigma_2(\omega, 1\text{ ps}) - \sigma_2^0(\omega)$ where $\sigma_2^0$ is the equilibrium conductivity. Gray dotted curve: change in imaginary conductivity $\Delta\sigma_2(\omega, \Delta T) = \sigma_2^0(\omega, 10\text{ K}) - \sigma_2^0(\omega, 60\text{ K})$ measured at equilibrium when cooling below Tc. **(a4)-(c4)** Light induced changes in reflectivity. Grey line: Change in reflectivity measured at negative

time delays (no signal). **(a5)-(c5)** Light induced changes in the electron loss function $\Delta(-\Im m\,1/\tilde{\varepsilon})$.

**Figure 6**

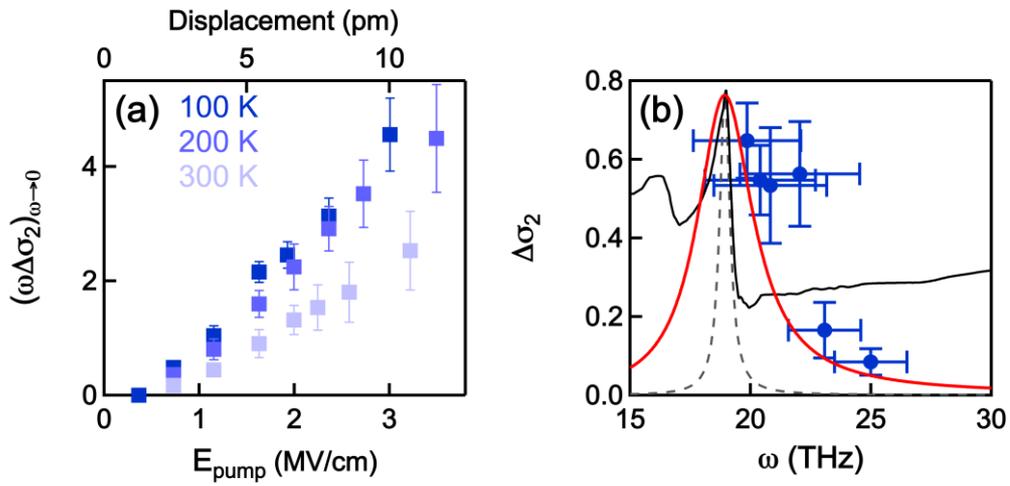

**Figure 6** Fluence and wavelength dependence. **(a)** Size of the photoinduced enhancement $\omega \Delta \sigma_2$ for different pump fluences at 100 K, 200 K, and 300 K. The corresponding estimated displacement of the oxygen atoms (see supplementary material) is shown in the top axis. **(b)** $\Delta \sigma_2$ at 1 THz plotted as a function of pump wavelength (blue dots) and measured at constant pump fluence. The horizontal bars indicate the pump spectral linewidth. The black line indicates equilibrium $\sigma_1(\omega)$. The resonant frequency of the apical oxygen mode is indicated in the dashed black curve, and in the red curve when convolved with the pump pulse frequency bandwidth.

**Figure 7**

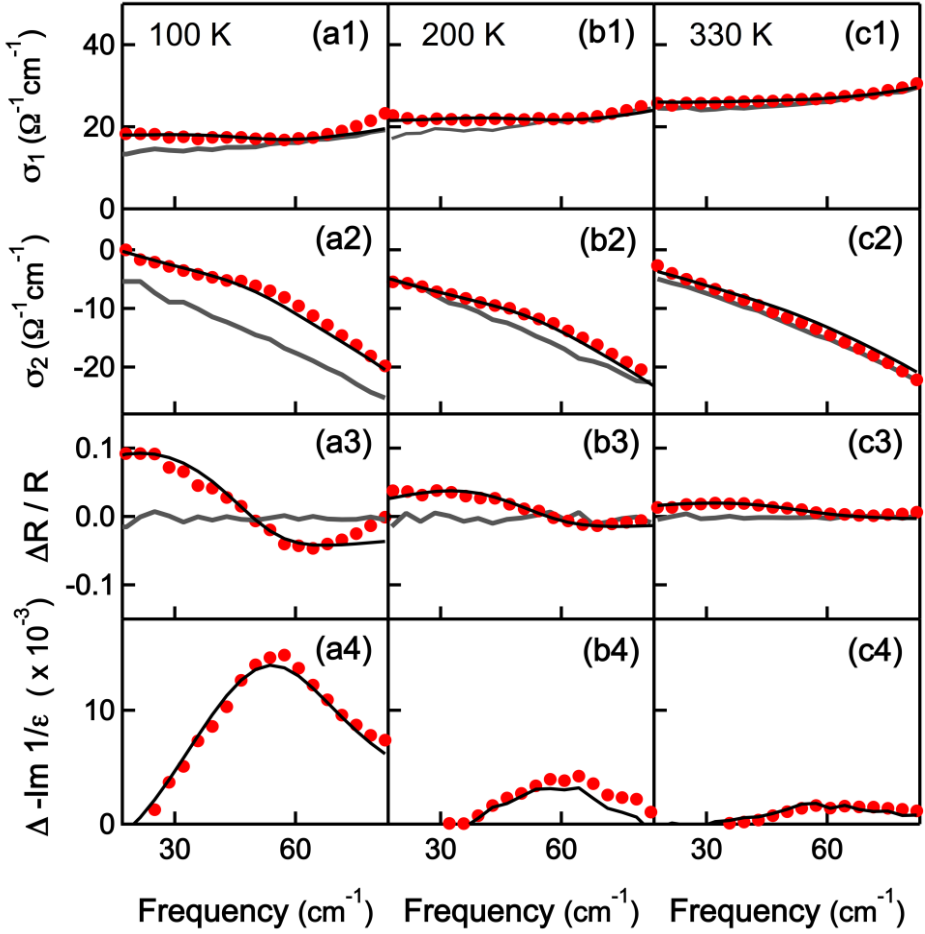

**Figure 7 Optical properties YBCO 6.5 above Tc** 0.8 ps time delay after excitation at (a) 100 K, (b) 200 K and (c) 330 K. Red dots: Light induced optical properties. Grey lines: properties of the equilibrium state. Black lines: effective medium fit for a transient conductor with $\tau_s$=7 ps. **(a1)-(c1)** Real part of the optical conductivity $\sigma_1(\omega)$. **(a2)-(c2)** Imaginary part of the optical conductivity $\sigma_2(\omega)$. **(a3)-(c3)** Light induced changes in reflectivity. Grey line: Change in reflectivity measured at negative time delays (no signal). **(a4)-(c4)** Light induced changes in the electron loss function $\Delta(-\Im m \, 1/\bar{\varepsilon})$.

**Figure 8**

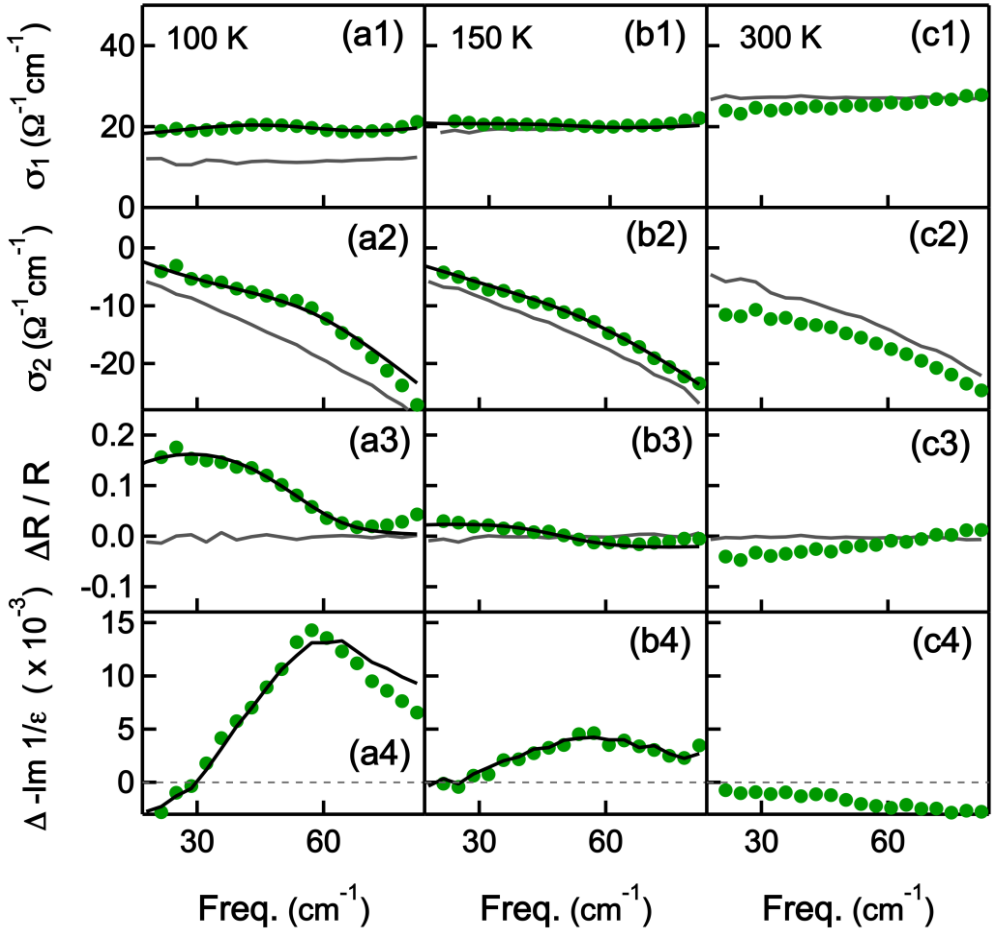

**Figure 8 Optical properties YBCO 6.6 above Tc.** 0.8 ps time delay after excitation at (a) 100 K, (b) 150 K and (c) 300 K. Green dots: Light induced optical properties. Grey lines: properties of the equilibrium state. Black lines: effective medium fit for a transient conductor with $\tau_s$=7 ps. **(a1)-(c1)** Real part of the optical conductivity $\sigma_1(\omega)$. **(a2)-(c2)** Imaginary part of the optical conductivity $\sigma_2(\omega)$. **(a3)-(c3)** Light induced changes in reflectivity. Grey line: Change in reflectivity measured at negative time delays (no signal). **(a4)-(c4)** Light induced changes in the electron loss function $\Delta(-\Im m\, 1/\bar{\varepsilon})$.

**Figure 9**

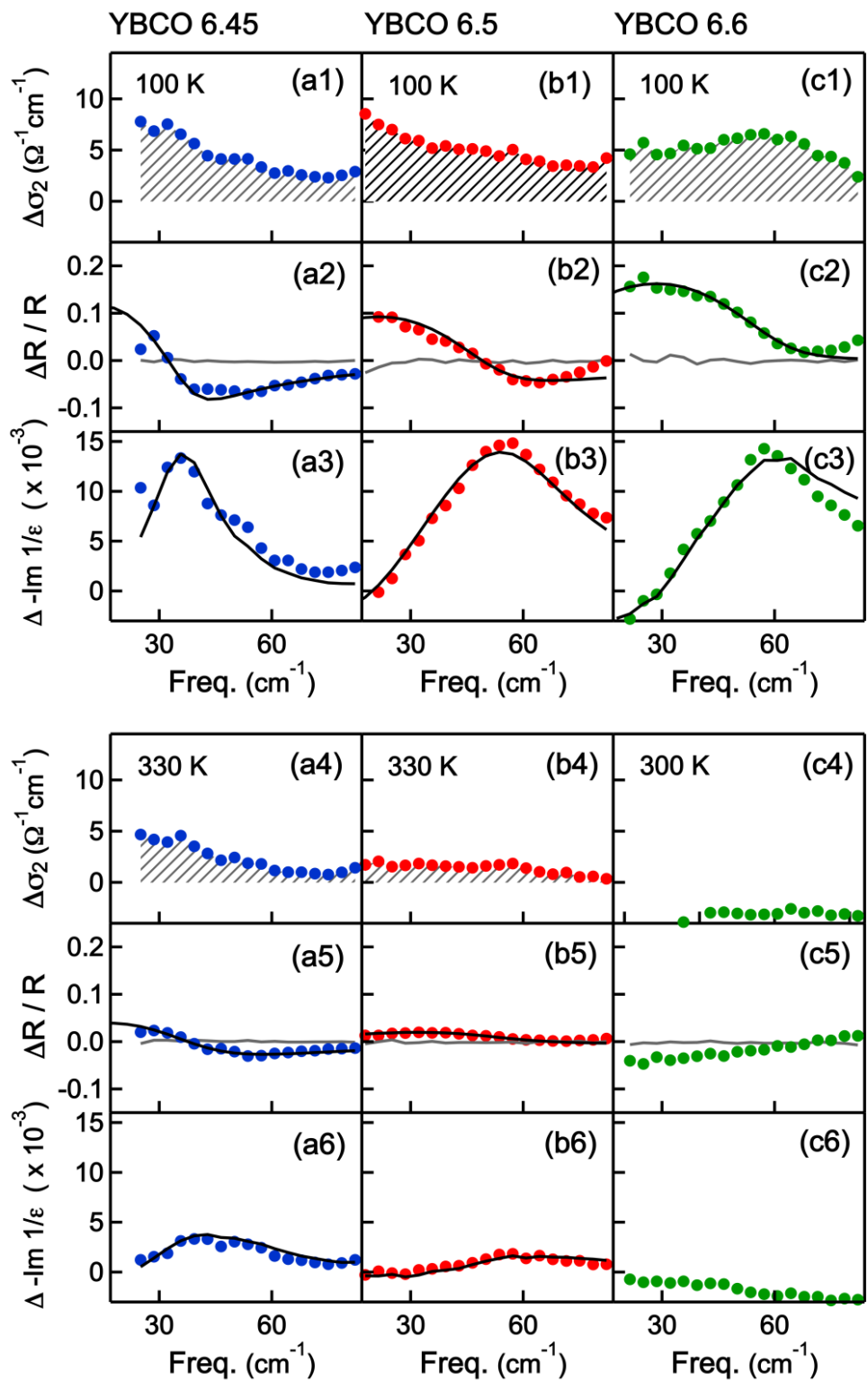

**Figure 9 Doping dependence transient optical properties** of **(a)** YBCO 6.45, **(b)** YBCO 6. 5, and **(c)** YBCO 6.6, 0.6 ps time delay after excitation (dots) at 100 K and high temperature. **(a1)-(c1)** Dots: Light induced change in the optical conductivity $\sigma_2(\omega)$ at 100 K. **(a2)-(c2)** Dots: Light induced changes in reflectivity at 100 K. Black line: effective medium fit. Grey line: Change in reflectivity measured at negative time delays (no signal). **(a3)-(c3)** Dots: Light induced

changes in the imaginary conductivity $\Delta\sigma_2(\omega)$ at 330 K (YBCO 6.45 and YBCO 6.5) and 300 K (YBCO 6.6). **(a5)-(c5).** Dots: Light induced changes in reflectivity at 330 K (YBCO 6.45 and YBCO 6.5) and 300 K (YBCO 6.6). Black line: effective medium fit. Grey line: Change in reflectivity measured at negative time delays (no signal). **(a6)-(c6).** Dots: Light induced changes in the electron loss function at 330 K (YBCO 6.45 and YBCO 6.5) and 300 K (YBCO 6.6). Black line: effective medium fit.

**Figure 10**

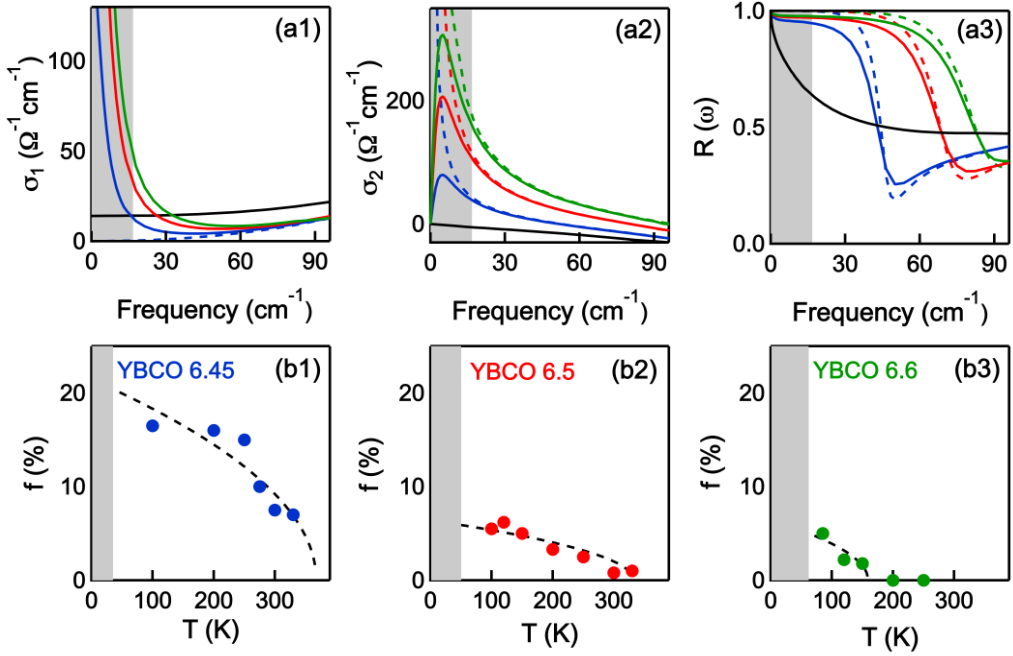

**Figure 10 Results of the effective medium analysis. (a1)-(a3).** Black curve: optical properties of the unperturbed material. Blue curves: extracted optical properties for the light induced phase in YBCO 6.45 at 100 K ($\sigma_1(\omega)$ in panel (a1), $\sigma_2(\omega)$ in panel (a2), reflectivity in panel (a3). Red Curves: extracted optical properties for the light induced phase in YBCO 6.5 at 100 K. Green curves: extracted optical properties for the light induced phase in YBCO 6.6 at 100 K. Solid lines assume a solid that exhibits perfect transport for the 7 ps lifetime of the state. Dashed curves represent the optical properties of a perfect conductor with infinitely long lifetime. Both assumptions provide a satisfactory fit in our spectral range. **(b1)-(b3).** Fitted filling fraction as a function of temperature. The dashed curves are fits obtained with an empirical mean-field dependence of the type $\propto \sqrt{1 - \frac{T}{T'}}$.

**Figure 11**

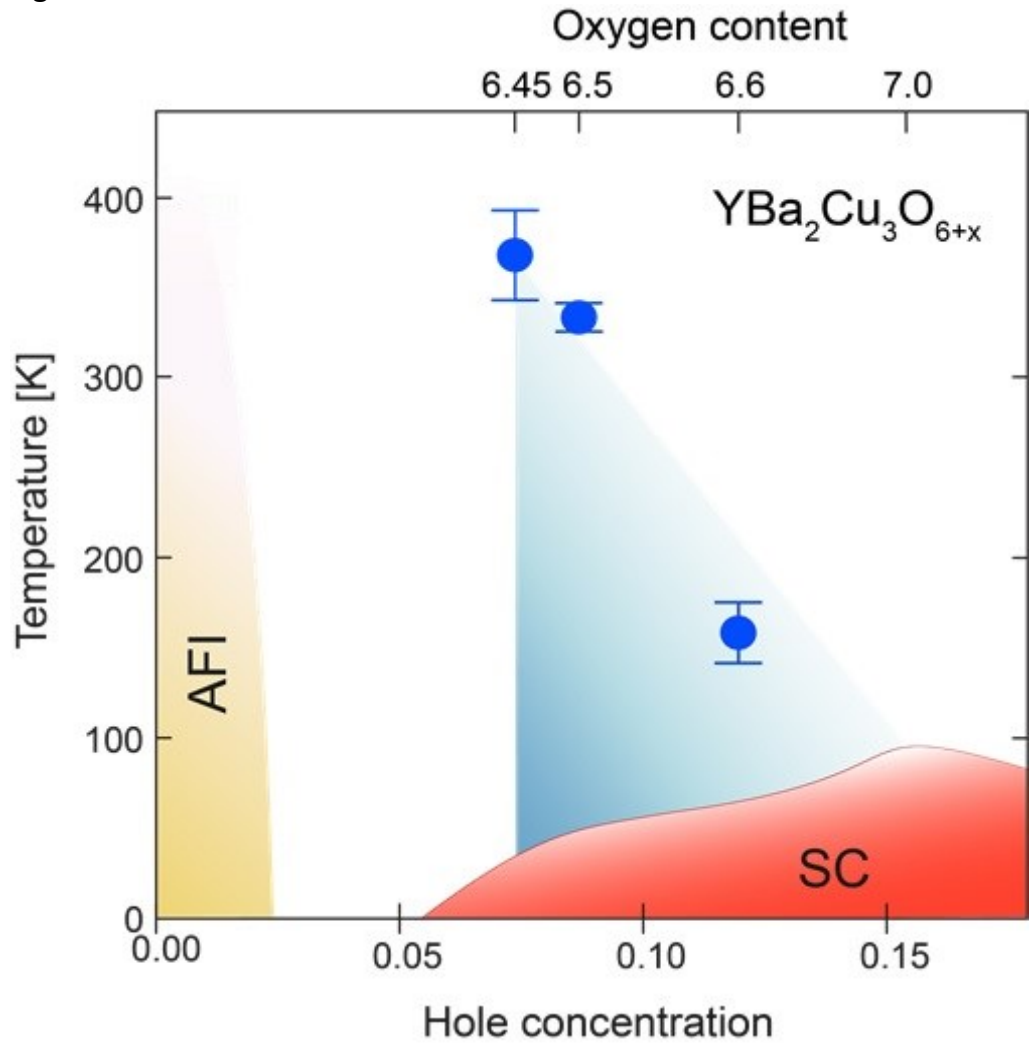

**Figure 11** Schematic phase diagram of YBCO, where the equilibrium phases (superconducting and antiferromagnetic), as well as the non-equilibrium phase discussed in this paper are displayed. The blue dots are the T' values extracted from the fits in figure 10(b1)-(b3).

# SUPPLEMENTARY INFORMATION

# Optically induced coherent transport far above T$_c$
# in underdoped YBa$_2$Cu$_3$O$_{6+\delta}$

S. Kaiser[1*], D. Nicoletti[1*], C.R. Hunt[1,4*], W. Hu[1], I. Gierz[1], H.Y. Liu[1], M. Le Tacon[2],

T. Loew[2], D. Haug[2], B. Keimer[2], and A. Cavalleri[1,3]

[1] Max Planck Institute for the Structure and Dynamics of Matter, Hamburg, Germany
[2] Max Planck Institute for Solid State Research, Stuttgart, Germany
[3] Department of Physics, Oxford University, Clarendon Laboratory, Oxford, United Kingdom
[4] Department of Physics, University of Illinois at Urbana-Champaign, Urbana, Illinois, USA

## S1 Sample Characterization

The $T_c$ values ($T_c$ = 35 K for YBCO 6.45, $T_c$ = 50 K for YBCO 6.5, $T_c$ = 62 K for YBCO 6.6, $T_c$ = 90 K for YBCO 7) were determined by dc magnetization measurements, as shown in Figure FS1(a). The measured superconducting transition was found to be sharp in all samples ($\Delta T_c \sim$2-4 K).

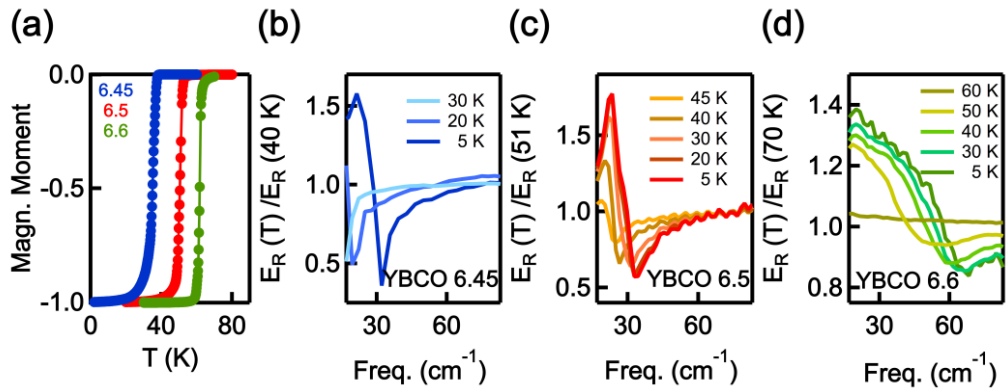

**Figure FS1 (a)** SQUID characterization of the dc magnetization across the superconducting transition. **(b)-(d)** Josephson plasma resonance in the relative THz reflected field ratios (below T$_c$ divided by above T$_c$) for YBCO 6.45, YBCO 6.5 and YBCO 6.6.

The equilibrium optical properties of the different crystals were characterized by THz time-domain spectroscopy. Broadband THz pulses (0.5-2.5 THz) were focused onto the sample surface at 30° incidence, with polarization perpendicular to the Cu-O planes. The reflected electric field was measured by electro-optic sampling at different temperatures, below and above $T_c$.

Figure FS1(b)-(d) display the absolute value of the frequency-dependent reflected electric field at a given temperature $|\tilde{E}_R(T)|$, normalized to the same quantity measured above $T_c$. Below $T_c$ a strong edge appears at a frequency $\omega_J$. This feature corresponds to the Josephson plasma resonance.

## S2 Derivation of the complex conductivity from differential reflectivity

The equilibrium electric field transient reflected by the sample $E_R(t)$ is measured by electro-optic sampling in ZnTe.

The pump-induced change in the reflected field was measured at each time delay $\tau$ during the dynamically evolving response of the material. For each pump-probe time delay $\tau$ the relative delay between excitation and sampling pulse was kept fixed, and the THz transient was scanned with respect to these two, changing the internal delay $t$. Therefore each point in the THz profile probed the material at the same time delay $\tau$ after excitation.

The complex reflection coefficient of the photo-excited sample, $\tilde{r}'(\omega, \tau)$, was determined from the normalized pump-induced changes to the electric field $\Delta\tilde{E}_R(\omega, \tau)/\tilde{E}_R(\omega)$ using the relation

$$\frac{\Delta\tilde{E}_R(\omega, \tau)}{\tilde{E}_R(\omega)} = \frac{\tilde{r}'(\omega, \tau) - \tilde{r}(\omega)}{\tilde{r}(\omega)}$$

where the stationary reflection coefficient $\tilde{r}(\omega)$ was evaluated from the equilibrium optical properties[i,ii]. The changes in the normal-incidence reflectivity were then calculated as $\frac{\Delta R}{R}(\omega, \tau) = (|\tilde{r}'(\omega, \tau)|^2 - |\tilde{r}(\omega)|^2)/|\tilde{r}(\omega)|^2$.

These "raw" reflectivity changes need to be reprocessed to take into account the mismatch between the several-μm penetration depth of the 0.5-2.5 THz probe and that of the resonant 20-THz pump, which is tuned to the middle of the *reststrahlen* band for this particular phonon and is evanescent over a few hundred nm (see Figure FS2(a)).

To extract the complex refractive index $\tilde{n}(\omega, \tau)$ of the photo-excited region, it was assumed that the change in $\tilde{n}(\omega, \tau)$ is maximum at the sample surface, decaying exponentially with distance, $z$, from the surface toward its unperturbed bulk value, $\tilde{n}_0(\omega)$, following Beer's law.

$$\tilde{n}(\omega, z) = \tilde{n}_0(\omega) + \Delta\tilde{n}(\omega)e^{-\alpha z},$$

where the linear extinction coefficient of the pump follows the inverse of the pump penetration depth, $\alpha = 1/d$.

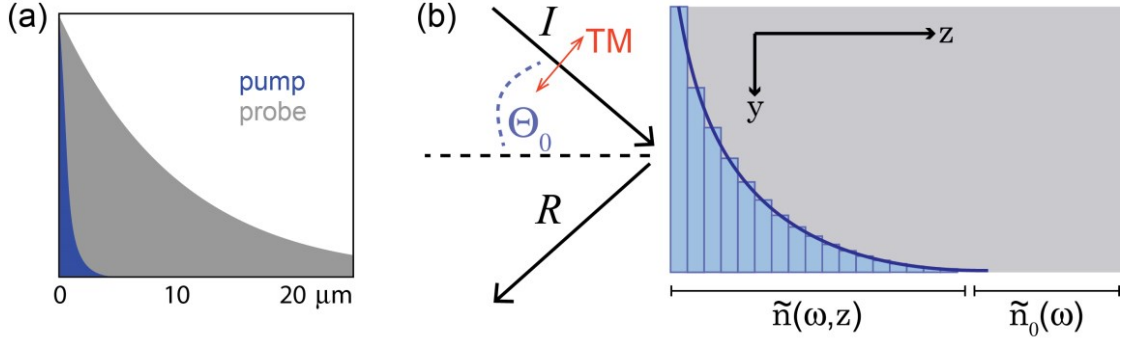

**Figure FS2** **(a)** Schematic of the penetration depth mismatch between resonant vibrational excitation (20-THz frequency or 15-μm wavelength, blue region) and the THz probe (0.5 – 2.5 THz, grey region) in YBCO 6.45. **(b)** The refractive index of the excited crystal surface is approximated as a series of thin layers of constant $\tilde{n}(z)$. From layer to layer, the refractive index decays exponentially to the bulk value, $\tilde{n}_0(\omega)$. The incidence angle in the experiment is $\theta_0 = 30°$.

To extract the optical properties of the excited medium we employ a transfer matrix formalism[iii]. We assume a stack of layers of vanishing thickness $\delta z$, where $\tilde{n}(z) = constant$ for each layer. For a single layer, $j$, the characteristic matrix reads:

$$M_j = \begin{pmatrix} \cos(k_j \tilde{n}_j \delta z) & -\dfrac{i}{p_j}\sin(k_j \tilde{n}_j \delta z) \\ -ip_j \sin(k_j \tilde{n}_j \delta z) & \cos(k_j \tilde{n}_j \delta z) \end{pmatrix},$$

with the refractive index $\tilde{n}_j = \tilde{n}(\omega, z = j\delta z)$, and $p_j = \cos\theta_j / \tilde{n}_j$ for a TM mode as in our experimental configuration. The THz probe is oriented $\theta_0 = 30°$ from normal incidence. The quantity $k_j$ is defined as $k_j = k_0 \cos\theta_j$, where the probe wave number in vacuum is $k_0 = \omega/c$. The total characteristic matrix, $M$, is obtained as the product of the matrices for all layers, $M = \prod_{j=0}^{N} M_j$. The thickness of each layer is set such that $\delta z \ll d$, and the layers range from the sample surface to the probe penetration depth, $L = N\delta z$. From the elements of the total characteristic matrix, $m_{ij}$, we can extract the reflection coefficient,

---

[iii] M. Born and E. Wolf, "Principles of Optics," Pergamon Press.

$$\tilde{r}'(\omega) = \frac{(m_{11} + m_{12}p_L)p_0 - (m_{21} + m_{22}p_L)}{(m_{11} + m_{12}p_L)p_0 + (m_{21} + m_{22}p_L)}.$$

The quantity $p_L$ is evaluated at the probe penetration depth and $p_0 = \cos\theta_0$ is calculated at the sample surface.

This equation is solved numerically for the surface refractive index $\tilde{n}(\omega, \tau) = \tilde{n}_0(\omega) + \Delta\tilde{n}(\omega, \tau)$ using the Levenberg-Marquardt fitting algorithm in Matlab.

From the surface refractive index, we calculate the complex conductivity for a volume that is homogeneously transformed,

$$\tilde{\sigma}(\omega, \tau) = \frac{\omega}{4\pi i}[\tilde{n}(\omega, \tau)^2 - \varepsilon_\infty],$$

where $\varepsilon_\infty = 4.5$, a standard value for cuprates[iv].

In the main text the changes in the reflectivity, $\frac{\Delta R}{R}(\omega, \tau) = \left(R(\omega, \tau) - R_{eq}(\omega)\right)/R_{eq}(\omega)$, are recalculated by assuming normal-incidence reflection,

$$R(\omega) = \left|\frac{1 - \tilde{n}(\omega)}{1 + \tilde{n}(\omega)}\right|^2.$$

In the limit $d \ll L$ — which is relevant for the full spectral range, at all temperatures, considered in this paper — we find the multilayer model introduced above in full agreement with a single-layer approximation in which we consider only one excited layer, $N = 1$, with thickness $\delta z = d$ on an unperturbed bulk.

### S3 Pump fluence dependence

MIR pumping transforms underdoped YBCO into an inhomogeneous coherent conductor, with an optical response consistent with perfectly conducting regions embedded in a bulk that retains its equilibrium optical properties.

The transient state of YBCO 6.45 was also measured as a function of pump fluence, as shown in Figure FS3. The photoinduced effect is present even at pump field strengths as low as 0.3 MV/cm (panels (a1)-(a3)), evidenced by a ~0.3% reflectivity edge.

Remarkably, the position of the edge exhibits no fluence dependence, thus excluding an interpretation of the transient conductivity in terms of a gas of photo-carriers with long scattering time.

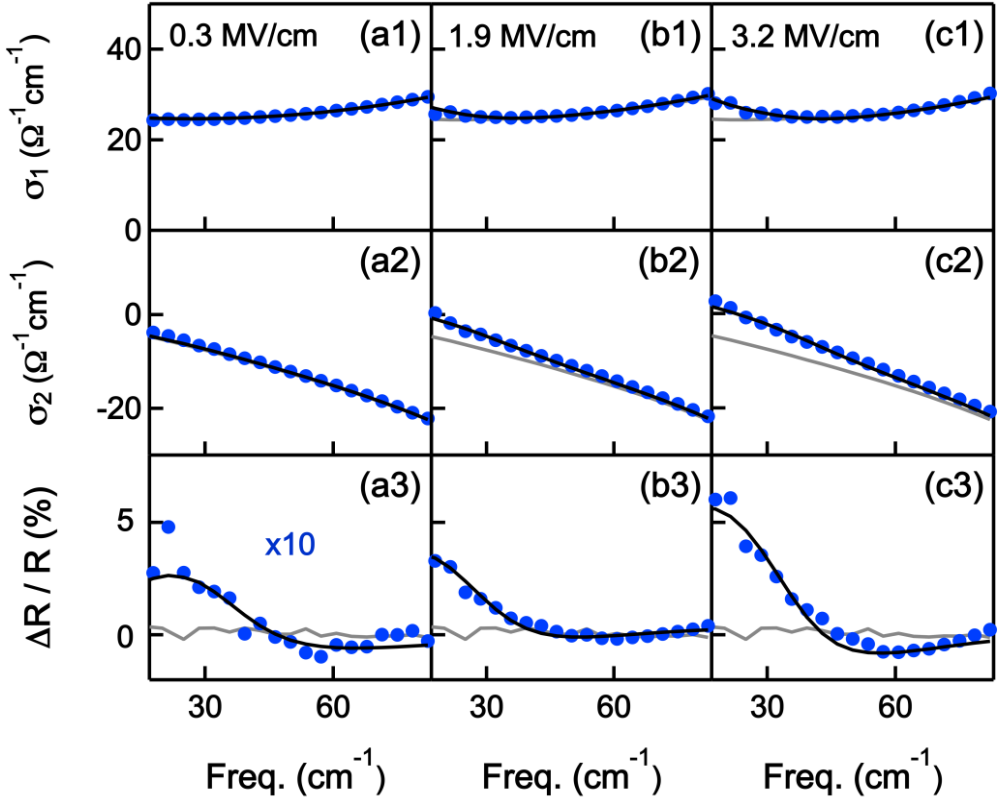

**Figure FS3**: Transient optical properties of YBCO 6.45 (blue dots) at 300 K at three different pump field strength values: **(a)** 0.3 MV/cm, **(b)** 1.9 MV/cm, and **(c)** 3.2 MV/cm. Black lines are effective medium fits to the photoinduced spectra, grey lines are the equilibrium optical properties. A distinct reflectivity edge at ~1 THz is observed in the transient state even at the lowest pump fluence.

## S4 Evaluation of the peak lattice distortion

For the electric fields of the pump pulse, the amplitude of the field-induced atomic displacement due to mid-infrared photo-excitation can be estimated assuming an ionic bonding between the apical $O^{2-}$ ion and the Cu(1) ion in the CuO chains.

The atomic polarizability can be derived as $\boldsymbol{P}(\omega_0) = \varepsilon_0 \chi(\omega_0) \boldsymbol{E}(\omega_0)$, where $\omega_0 = 20$ THz and $\|\boldsymbol{E}(\omega_0)\| \cong 3$ MV/cm. The susceptibility $\chi(\omega_0)$ is calculated from the $c$-axis equilibrium optical conductivity using $|\varepsilon_0 \chi(\omega_0)| = \left| \frac{\sigma(\omega_0)}{\omega_0} \right|$. As $|\sigma_1(\omega_0)| \cong |\sigma_2(\omega_0)| \cong$

$30 \, \Omega^{-1}\text{cm}^{-1}$ (from data in Refs. X and Y), one gets $|\varepsilon_0 \chi(\omega_0)| \cong 2 \cdot 10^{-12} \Omega^{-1}\text{cm}^{-1}\text{s}$ and $\|\boldsymbol{P}(\omega_0)\| \cong 6 \cdot 10^{-6} \, \text{C} \cdot \text{cm}^{-2}$.

The average size of the photo-induced electric dipole, i.e., the displacement of the oxygen ions, is then given by $d = \|\boldsymbol{P}(\omega_0)\|/nQ$, where $n$ is the density of dipoles (2 per unit cell of volume 173 $\text{Å}^3$) and $Q = 3e$. This yields $d \sim 10$ pm, which is approximately 5% of the equilibrium Cu-O distance.

## References (main text):

YBCO 6.5 and 6.6, an additional incoherent component was added to $\tilde{\varepsilon}_{NS}(\omega)$ to account for the quasi-particle contribution to the transient state: This can be seen most clearly in YBCO 6.6 as a flat increase in the ohmic conductivity (figure 8.a1). This contribution is negligible in YBCO 6.45 up to 100 K, but becomes finite at higher temperatures and, especially, as one approaches optimal doping. This quasi-particle component evolves following the short decay timescale and can be modeled in our bandwidth range as a single parameter: a frequency-independent contribution to $\sigma_1(\omega)$ that affects only the non-transformed volume $1-\square$.